\documentclass[12pt]{spieman}  
\usepackage{amsmath,amsfonts,amssymb}
\usepackage{graphicx}
\usepackage{setspace}
\usepackage{tocloft}

\title{Synchronization and temporal nonreciprocity of optical microresonators via spontaneous symmetry breaking}

\author[a]{Da Xu}
\author[a]{Zi-Zhao Han}
\author[a]{Yu-Kun Lu}
\author[a,b,c,d]{Qihuang Gong}
\author[e]{Cheng-Wei Qiu}
\author[c,f,*]{Gang Chen}
\author[a,b,c,d,*]{Yun-Feng Xiao}
\affil[a]{State Key Laboratory for Artificial Microstructure and Mesoscopic Physics, School of Physics, Peking University, Beijing 100871, China}
\affil[b]{Nano-optoelectronics Frontier Center of the Ministry of Education (NFC-MOE)
Collaborative Innovation Center of Quantum Matter, Beijing 100871, China
}
\affil[c]{Collaborative Innovation Center of Extreme Optics, Shanxi University, Taiyuan 030006, People's Republic of China}
\affil[d]{Beijing Academy of Quantum Information Sciences, Beijing 100193, China}
\affil[e]{Department of Electrical and Computer Engineering, National University of Singapore,
Singapore 117576, Singapore}
\affil[f]{State Key Laboratory of Quantum Optics and Quantum Optics Devices, Institute of Laser Spectroscopy, Shanxi University, Taiyuan 030006, People's Republic of China}

\cftpagenumbersoff{figure}
\cftpagenumbersoff{table} 
\begin{document} 
\maketitle
\begin{abstract}
Synchronization is of importance in both fundamental and applied physics, but their demonstration at the micro/nanoscale is mainly limited to low-frequency oscillations like mechanical resonators.
Here, we report the synchronization of two coupled optical microresonators, in which the high-frequency resonances in optical domain are aligned with reduced noise.
It is found that two types of synchronization emerge with either the first- or second-order transition, both presenting a process of spontaneous symmetry breaking.
In the second-order regime, the synchronization happens with an invariant topological character number and a larger detuning than that of the first-order case.
Furthermore, an unconventional hysteresis behavior is revealed for a time-dependent coupling strength, breaking the static limitation and the temporal reciprocity.
The synchronization of optical microresonators offers great potential in reconfigurable simulations of many-body physics and scalable photonic devices on a chip.
\end{abstract}


{\noindent \footnotesize\textbf{*}Yun-Feng Xiao,  \linkable{yfxiao@pku.edu.cn}; }
{ \footnotesize Gang Chen,  \linkable{chengang@sxu.edu.cn} }
\begin{spacing}{2}   
The phenomena of synchronization are ubiquitously observed in nature like the collective neuron bursts, the stabilized heartbeats, and the disciplined synchronous fireflies \cite{van1928heart,fitzhugh1961neuronimpulses,buck1968fireflymechanism}. Starting from the Huygens pendulum locked in anti-phase \cite{huygens1897oeuvres,oliveira2015huygens}, the synchronization of nonlinear oscillators has earned in-depth investigation \cite{kurths2001synchronization}.  
In the daily life and modern industries, the synchronization has been the basis for clock calibration, signal processing, and microwave communication \cite{bregni2002clock}, and provides novel schemes of clustered computing and memory storage \cite{bagheri2011dynamic,mahboob2008bit,hoppensteadt2001synchronization_neurocom}.
Over the past few years, the synchronization of mechanical resonators has been implemented, where the mechanical resonators are coupled strongly through direct conjunction elements \cite{nijmeijer2003synchronization,shim2007synchronized}, optical radiation fields \cite{Heinrich2011collective,holmes2012sync,zhang2012synchronization,zhang2015synchronization,peano2015topological} or optical traveling waves \cite{2015master,bagheri2013photonic,li2016oe,2017locking}, facilitating the mechanical-based high performance networks. The strong mutual coupling together with the nonlinearity of individually sustainable systems plays a crucial role in realization of synchronization \cite{cross2004synchronization,Agrawal2013locked,walter2014quantum,pecora2014cluster,matheny2014phase,lorch2016genuine}.

Likewise, synchronized optical fields shall also promise great potentials in fundamental and applied physics, such as many-body optical physics and scalable on-chip photonic devices \cite{kuramochi2014large,zhang2019electronically,liu2014coherentQED,hwang2016quantum,tangpanitanon2016topological,ludwig2013quantum}, while the occurrence is challenged by their relatively low mutual coupling compared to the high carrier frequencies of light.
Recently the microcomb solitons are synchronized experimentally \cite{jang2018synchronization,yang2017counter}, significantly expanding their photonic applications, yet the repetition rates in the range of microwaves rather than the optical frequency of the comb lines are equalized.
In this article, we study the mode synchronization of two optical microresonators without an external reference frequency, where the distant modes are self-sustained and mutually aligned through a weak coupling.
The synchronization results from the spontaneous symmetry breaking and takes the form of a first- or second-order transition.
Furthermore, an unconventional hysteresis behavior is presented as the coupling strength varies, permitting the nonreciprocal synchronization in a more extensive parametric space.

\begin{figure}[t]
\centering
\includegraphics[width=12cm]{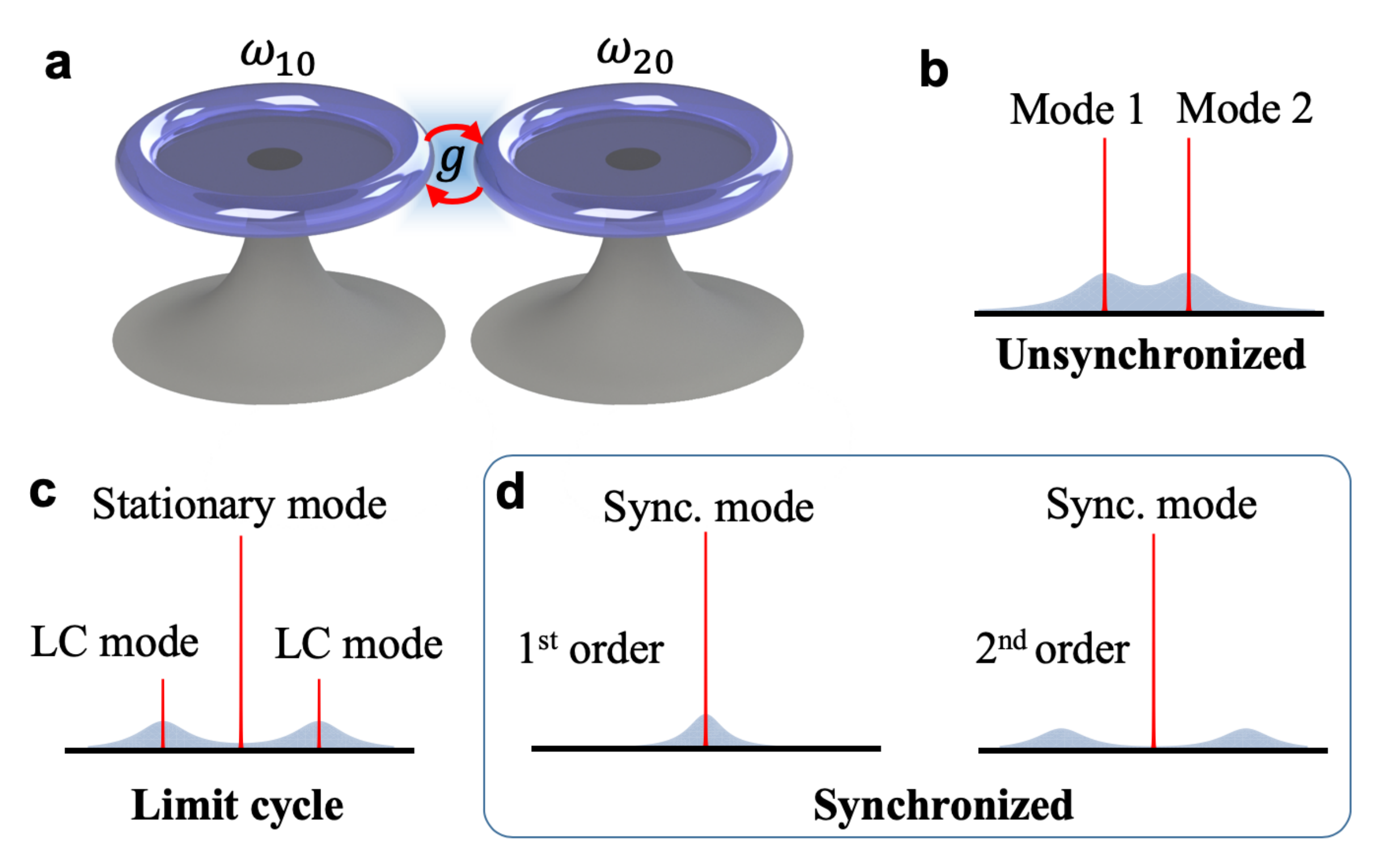}
\caption{\textbf{Schematic diagram of the system.}
\textbf{a}, two detuned and self-sustained optical microcavities with different resonant frequencies $\omega_{10}$ and $\omega_{20}$, which are directly coupled at a strength $g$.
\textbf{b-d}, frequency spectra of the coupled cavities, showing three different long-term states: the unsynchronized, limit cycle (LC) and synchronized (Sync.) states. Light blue presents noise backgrounds, from which the first- and second-order synchronizations are distinguished.}
\label{setup}
\end{figure}

\begin{figure*}[!h]
\centering
\includegraphics[width=17cm]{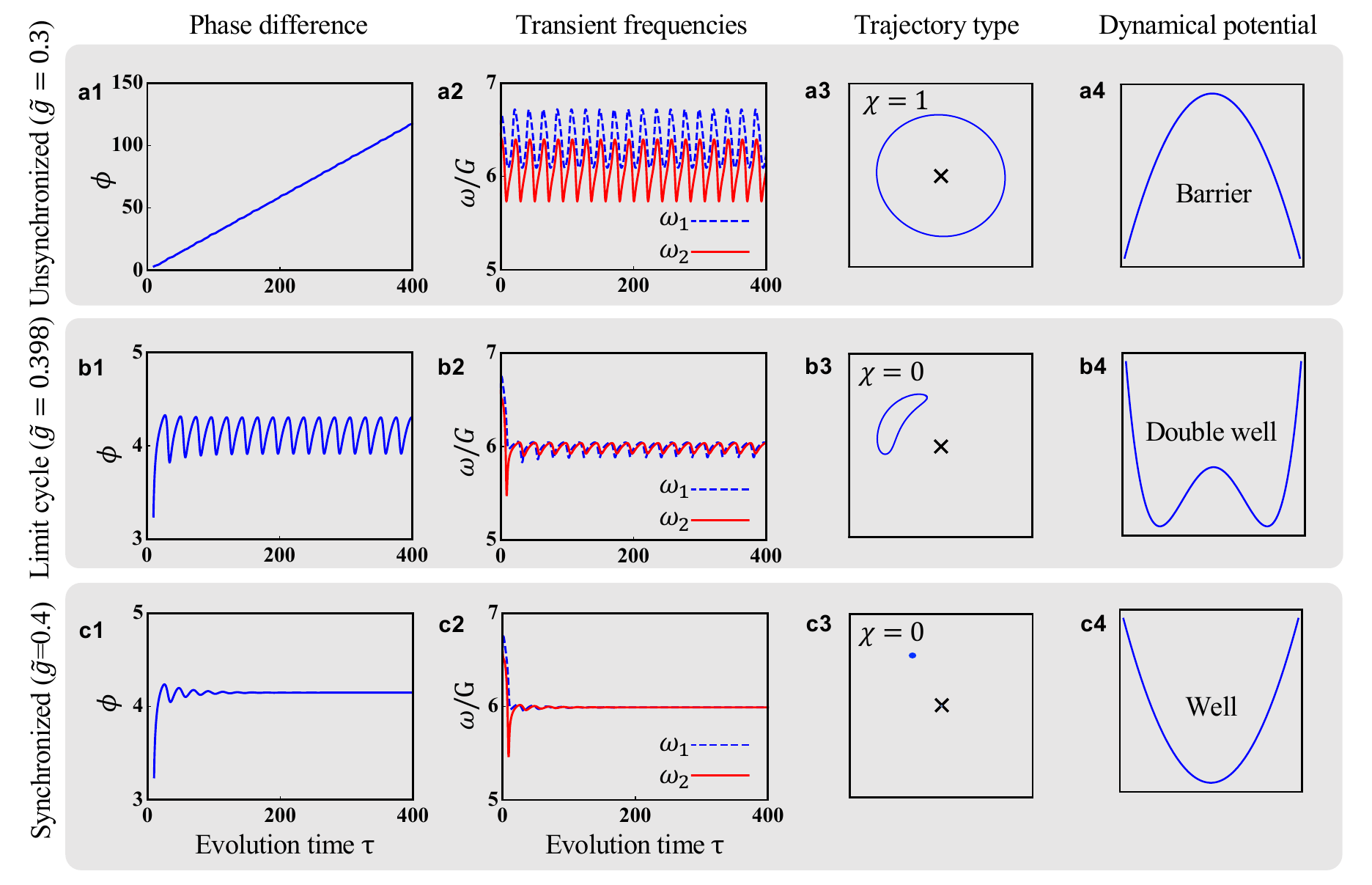}
\caption{\textbf{Long-term evolutions of the two cavity modes under different coupling strengths.}
Three different categories are shown: the unsynchronized ($\tilde g=0.3 $), limit cycle  ($\tilde g=0.398$) and synchronized states ($\tilde g=0.4$) in \textbf{a}, \textbf{b} and \textbf{c}.
\textbf{a1-c1}, phase difference; 
\textbf{a2-c2}, transient frequencies;
\textbf{a3-c3}, trajectory encircling types (black cross as the axis);
\textbf{a4-c4}, dynamical potential near the synchrony point. In all subfigures, the given detuning $\tilde \Delta=0.3$ and Kerr factor $\tilde \delta=0.1$. 
}
\label{diffcoup1}
\end{figure*}

\section*{Results}
\noindent \textbf{Two self-sustained microresonators and interaction model.}
As shown in Fig.~\ref{setup}\textbf{a}, the system is composed of two optical microcavities with different resonant frequencies $\omega_{10}$ and $\omega_{20}$, coupled at the strength $g$. The $j$th ($j=1,2$) cavity is self-sustained by the internal gain described by the factor $G_{\text{p},j}$ and the intrinsic dissipation at the rate $\kappa_j$. In the presence of nonlinear gain, an self-Kerr type  modulation $\delta_j (a_{j}^{\dagger} a_j)^2/2$ is present, $a_j$ being the annihilation operator and the factor $\delta_j$ describing the self-Kerr effect \cite{jirauschek2006gaussian}.
With the gain saturation \cite{liu2017symmetry} or the multi-photon absorption \cite{lee2013quantum,walls2007quantum,gilles1993two}, the effective dissipation of the $j$th mode is modeled as $K_j=\kappa_j/2+R_j\langle a_{j}^{\dagger} a_j\rangle$, where $R_j$ is the nonlinearity factor.

The dissipative evolution of the system is described by the Lindblad density-matrix equation ($\hbar=1$ hereafter),
\begin{eqnarray}
\dot\rho=-i[H,\rho]+\sum_{j=1,2}\left(G_j\mathcal{D}[a_j^\dagger]\rho+\frac{R_j}{2}\mathcal{D}[a_j^2]\rho\right).\label{lindblad}
\end{eqnarray}
Here $\mathcal{D}[o]\rho=2o\rho o^\dagger-o^\dagger o\rho-\rho o^\dagger o$ and $G_j=G_{\text{p},j}-\kappa_j/2$ denotes the net gain factor. Without an external frequency reference, the time-independent Hamiltonian 
$H=\sum_{j=1,2}[\omega_{j0}a_j^{\dagger}a_j
+\delta_j(a_j^{\dagger}a_j)^2/2]+g(a_2^{\dagger}a_1+a_1^{\dagger}a_2)$, under the rotating-wave approximation.
For the simplicity, in the following  we set $R_1=R_2=R$, $G_1=G_2=G$ and $\delta_1=\delta_2=\delta$, and the dimensionless parameters are defined as $\tilde\delta=\delta/R$, $\tilde\Delta=(\omega_{10}-\omega_{20})/G$, $\tilde g=g/G$ and the time scale $\tau= Gt$. These formalism can be checked from wave functions in systems like coupled laser systems\cite{sebastian2004bifur,wieczorek2004chaos}.
Though the coupling between the two modes is linear and energy conservative, it plays the role of messenger passing over the weak and detuned drive. The self-sustained system always favors the resonance mutual driving, and the synchronization of two modes is established by the spontaneous frequency alignment of the individual cavities, through the self-Kerr effect and the amplitude stabilization under saturation effect. 
In this way, the modes are synchronized in individual cavities.
\vspace{6pt}

\noindent \textbf{Synchrony solution in static case.}
We focus on the phase difference and the transient frequencies of two modes in the coherent-state representation \cite{kurths2001synchronization}.
In this representation, the complex amplitude $\alpha_j=\langle a_j\rangle$ is parameterized as $r_j\sqrt{G/R}e^{-i\phi_j}$, where $r_j$ and $\phi_j$ are the amplitude and the phase, respectively. Let $\phi=\phi_1-\phi_2$ be the phase difference, which is the preserved degree of freedom, and $\omega_j=\dot\phi_j$ be the transient frequency for the $j$th mode.

Following the standard Wigner function formalism \cite{carmichael2000statisticaloptics}, the mode equation is described by $(\mathbf{\dot\Lambda},\mathbf{\bar{\dot\Lambda}})^\intercal=\mathbf{f}(\mathbf{\Lambda},\mathbf{\bar\Lambda})$, where $\mathbf{\Lambda}=(\alpha_1,\alpha_2)$, and $\mathbf{f}(\mathbf{\Lambda},\mathbf{\bar\Lambda})$ denotes the quasi-probability drift flow of two modes \cite{supp}. The synchrony solution is achieved when $\mathbf{f}(\mathbf{\Lambda},\mathbf{\bar\Lambda})=0$, and a fixed point $\mathbf{\Lambda}_\text{s}(\tilde\Delta,\tilde\delta,\tilde g)$ emerges in the parametric space \cite{supp}.
In Fig.~\ref{diffcoup1} we plot the phase differences and the transient frequencies for different $\tilde{g}$. Three categories of long-term behaviors are discovered. When the coupling strength is weak, $\tilde{g}=0.3$ for example, the phase difference $\phi$ accumulates to infinity quickly [see Fig.~\ref{diffcoup1}\textbf{a1}], and the transient frequencies $\omega_1$ and $\omega_2$ are effectively separated [Fig.~\ref{diffcoup1}\textbf{a2}], showing two separated modes in the frequency spectrum [Fig.~\ref{setup}\textbf{b}].
When the coupling strength is turned higher like $\tilde{g}=0.398$, the phase difference $\phi$ vibrates around the stationary point but does not accumulate [Fig.~\ref{diffcoup1}\textbf{b1}], and the frequencies $\omega_1$ and $\omega_2$ breath slowly around the stationary frequency [Fig.~\ref{diffcoup1}\textbf{b2}], generating a limit cycle state. A stationary mode is localized around the original two cavity modes, while a pair of weak limit cycle modes could be found symmetrically detuned from the stationary modes [Fig.~\ref{setup}\textbf{c}].
Finally, with a strong enough coupling strength like $\tilde{g}=0.4$, the phase difference $\phi$ stabilizes [Fig.~\ref{diffcoup1}\textbf{c1}], and the frequencies $\omega_1$ and $\omega_2$ also converge to a single value [Fig.~\ref{diffcoup1}\textbf{c2}], reaching the synchronized state with a single mode in the frequency spectrum [Fig.~\ref{setup}\textbf{d}].

It is noted that the temporal translational symmetry (TTS) is preserved in the synchronized state because both the amplitudes and phase difference remain invariant, while the symmetry is broken in the unsynchronized and limit cycle states.
The discrete topological character number as the average encircling number is further defined
\begin{eqnarray}
\chi=\frac{T_0}{2\pi}\lim_{T\to\infty}\left|\frac{1}{T}\int_{0}^{T}dt(\omega_1-\omega_2)\right|,
\end{eqnarray}
with $T_0$ being the period of long-term evolution \cite{mari2013measures,supp}. 
As shown in Fig.~\ref{diffcoup1}\textbf{a3}, the unsynchronized trajectory encircles the axis $r_1=r_2=0$ and has the character number $\chi=1$. The later two categories of trajectories, the off-axial circles [Fig.~\ref{diffcoup1}\textbf{b3}] and the fixed points [Fig.~\ref{diffcoup1}\textbf{c3}], have the character number $\chi=0$ [see detailed transformed space in \cite{supp}].
With the different symmetries and character numbers, the three long-term states are classified accordingly \cite{supp}.
\vspace{6pt}

\begin{figure}[!h]
\centering
\includegraphics[width=12cm]{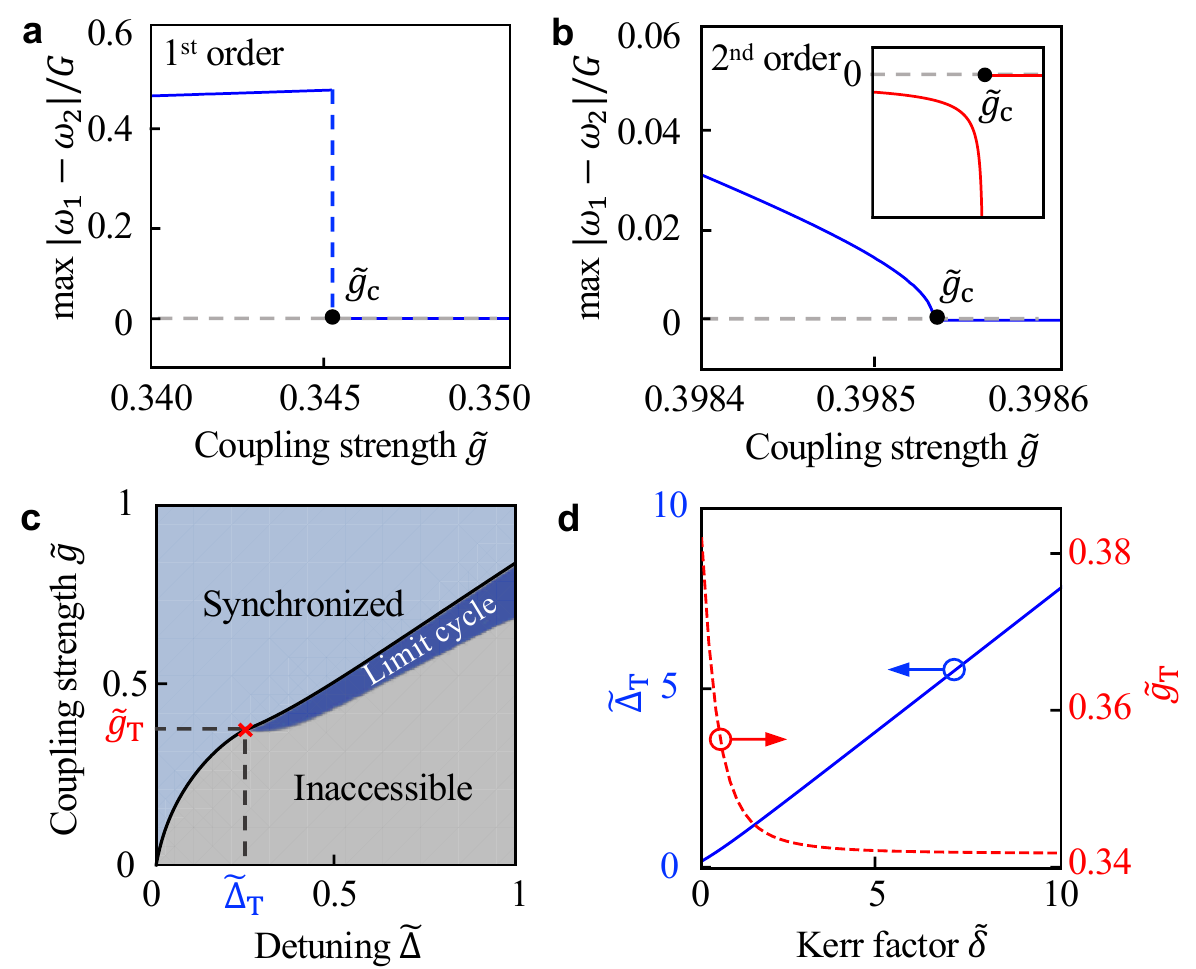}
\caption{\textbf{Parameter dependence of the synchronization. }Maximum of the frequency differences, $\max |\omega_1 -\omega_2|$, versus the coupling strength $\tilde g$,
with ($\tilde\Delta=0.2$, $\tilde\delta=0.1$) in \textbf{a} and ($\tilde\Delta=0.3$, $\tilde\delta=0.1$) in \textbf{b}. Inset shows the derivative. \textbf{c,} Phase diagram in the $(\tilde\Delta,\tilde g)$ plane with the Kerr factor $\tilde\delta=0.1$.
The inaccessible (grey), limit cycle (dark blue) and synchronized (light blue) regimes are marked. 
The red cross stands for the triple phase point $(\tilde\Delta_\text{T},\tilde g_\text{T})$.
\textbf{d,} The triple phase point $(\tilde\Delta_\text{T},\tilde g_\text{T})$ depending on the Kerr factor $\tilde\delta$. 
}
\label{surface_g}
\end{figure}

\noindent \textbf{Analysis on different transition types.}
We further study the maximum of the frequency differences, $\text{max} |\omega_1-\omega_2|$,
and two types of synchronization transitions are found.
In Fig.~\ref{surface_g}\textbf{a}, when the coupling strength $\tilde g$ is weak, the maximal frequency difference varies slowly. At a critical strength $\tilde{g}_\text{c}$, it suddenly falls to zero, which shows the characteristics of the first-order transition. 
In Fig.~\ref{surface_g}\textbf{b}, the maximal frequency difference continuously decreases to zero but has a discontinuity in its derivative at $\tilde g_\text{c}$, showing the feature of the second-order transition. 
Besides, the noise spectrum is also calculated in long-term motions. For synchronized spectrum in Fig.~\ref{setup}\textbf{d}, the background noise has coinciding peaks with the synchronized frequencies in the first-order transition, while the noise has shifted-away peaks in the second-order transition \cite{supp}. 

In order to study the critical coupling strength $\tilde g_\text{c}$ and the transition behaviors in its vicinity, a real-valued dynamical potential $V(\mathbf{\Lambda},\mathbf{\bar\Lambda})$ is defined \cite{supp}. 
Only if the dynamical potential has a local minimum, the fixed point $\mathbf{\Lambda}_\text{s}(\tilde\Delta,\tilde\delta,\tilde g)$ emerges and remains stable, and thus indicates the existence of a synchronized state \cite{supp}.
In the vicinity of the fixed point, the dynamical potential can be expanded as
$V(\mathbf{\Lambda},\mathbf{\bar\Lambda})=V(\mathbf{\Lambda}_\text{s},\mathbf{\bar\Lambda}_\text{s})\notag-\!\frac{1}{2}\!\left[(\Delta\mathbf{\bar\Lambda},\Delta\mathbf{\Lambda})\!\cdot\!\mathbf{J}\!\cdot\!(\Delta\mathbf{\bar\Lambda},\Delta\mathbf{\Lambda})^\dagger+\text{H.c.}\right]\label{expansion}$,
where $\mathbf{J}(\mathbf{\Lambda},\mathbf{\bar\Lambda})=\partial \mathbf{f}(\mathbf{\Lambda},\mathbf{\bar\Lambda})/\partial(\mathbf{\Lambda},\mathbf{\bar\Lambda})$
is the Jacobian matrix, $\Delta\mathbf{\Lambda}$ is the arbitrarily small displacement from the fixed point, and H.c. is the Hermitian conjugate. The displacement $\Delta\mathbf{\Lambda}$ signifies the breaking of the TTS. At $\Delta\mathbf{\Lambda}=0$, the TTS is preserved. The stability near the fixed point is thus governed by the eigenvalues of the Jacobian.  
When the largest real part of the $\mathbf{J}$ eigenvalues [known as the largest Lyapunov exponent $\mathcal{L}(\mathbf{\Lambda}_\text{s})$] is positive, the fixed point is unstable and vice versa \cite{supp}. The critical coupling strength $\tilde{g}_\text{c}$ is then taken at $\mathcal{L}(\mathbf{\Lambda}_\text{s})=0$.

In the three-dimensional space ($r_1,r_2,\phi$), the Jacobian $\mathbf{J}$ has purely real $3\times3$ components, and thus the complex eigenvalues must come in pairs. 
If the largest Lyapunov exponent $\mathcal{L}$ equals one of the eigenvalues, the dynamical potential is simplified as \cite{supp}
\begin{eqnarray}
V(x)=b_0(\tilde g-\tilde g_\text{c})x^2\label{singlewell}
,\end{eqnarray} 
where $x$ is the perturbation of $\mathbf{\Lambda}_\text{s}$ in the direction of corresponding eigenvector, and the real coefficient $b_0=-d\mathcal{L}/d\tilde{g}>0$.
It is noted that the dynamical potential in Eq.~(\ref{singlewell}) becomes a well or barrier depending on $\tilde g>\tilde g_\text{c}$ or $\tilde g<\tilde g_\text{c}$, which leads to the synchronized or unsynchronized state in Fig.~\ref{diffcoup1}\textbf{a4} and \ref{diffcoup1}\textbf{c4}.
Thus the first-order transition happens at $\tilde g_\text{c}$, explaining the sudden convergence of frequency difference in Fig.~\ref{surface_g}\textbf{a}. For $\tilde g <\tilde g_\text{c}$ and  $\tilde g >\tilde g_\text{c}$, the TTS is broken ($\Delta\mathbf{\Lambda}\to\infty$) and preserved ($\Delta\mathbf{\Lambda}=0$), respectively. If the largest Lyapunov exponent $\mathcal{L}$ equals the real parts of a pair of conjugating eigenvalues, the averaged dynamical potential 
\begin{eqnarray}
\langle V \rangle=b_1(\tilde g-\tilde{g}_\text{c})\rho^2+b_2\rho^4 \label{double_well},
\end{eqnarray} 
where $\rho$ is the radial displacement from $\mathbf{\Lambda}_\text{s}$ and the real coefficients $\{b_1,  b_2\}>0$ \cite{supp}. 
When $\tilde g<\tilde g_\text{c}$, a double-well type potential is obtained, corresponding to the limit cycle state in Fig.~\ref{diffcoup1}\textbf{b4}.
After $\tilde g$ surpasses $\tilde g_\text{c}$, the averaged dynamical potential $\langle V \rangle$ has a single local minimum at $\mathbf{\Lambda}_\text{s}$, and the synchronization is reached, accounting for the second-order transition in Fig.~\ref{surface_g}\textbf{b}. The TTS is spontaneously broken as the the radial displacement $\rho$ continuously departs from the synchrony point $\mathbf{\Lambda}_\text{s}$.

In the light of the static analysis above, the phase diagram in the $\tilde\delta$-cross section is plotted in Fig.~\ref{surface_g}\textbf{c}, where three regions of different long-term behaviors are marked.
The synchronized and limit cycle regimes are specified according to the existence of a single and double local minima of the dynamical potentials, respectively.
The inaccessible (unsynchronized) regime corresponds to the saddle nodes in the dynamical potentials, and thus neither the synchronized state nor the limit cycle state can survive in this regime. 
The transition from the unsynchronized state to the synchronized state is of first-order and has a variant topological character number. The transition from the limit cycle state to the synchronized state is of second-order and has an invariant topological character number \cite{supp}. It is also found that a triple phase point emerges at $\tilde g_\text{c}=\tilde g_\text{T}$ and $\tilde\Delta=\tilde\Delta_\text{T}$, where $\tilde\Delta_\text{T}$ is the minimal detuning required for the second-order transition.
This point corresponds to the solution where two eigenvalues of the Jacobi matrix $\mathbf{J}$ are zeros. 
For the detuning $\tilde \Delta<\tilde\Delta_\text{T}$ ($\tilde \Delta>\tilde\Delta_\text{T}$), the first-order (second-order) transition happens around $\tilde g_\text{c}$ (black solid line).
The triple phase point relies crucially on the strength of the Kerr effect. In Fig.~\ref{surface_g}\textbf{d}, we plot $\tilde\Delta_\text{T}$ and $\tilde g_\text{T}$ with respect to the Kerr factor $\tilde \delta$, showing monotone increasing and decreasing dependence, respectively.
When the factor $\tilde\delta$ increases, the self-tuning ability of the Kerr effect is strengthened, and the second-order synchronization under a larger detuning and a weaker coupling becomes possible.

\begin{figure}
\centering
\includegraphics[width=12cm]{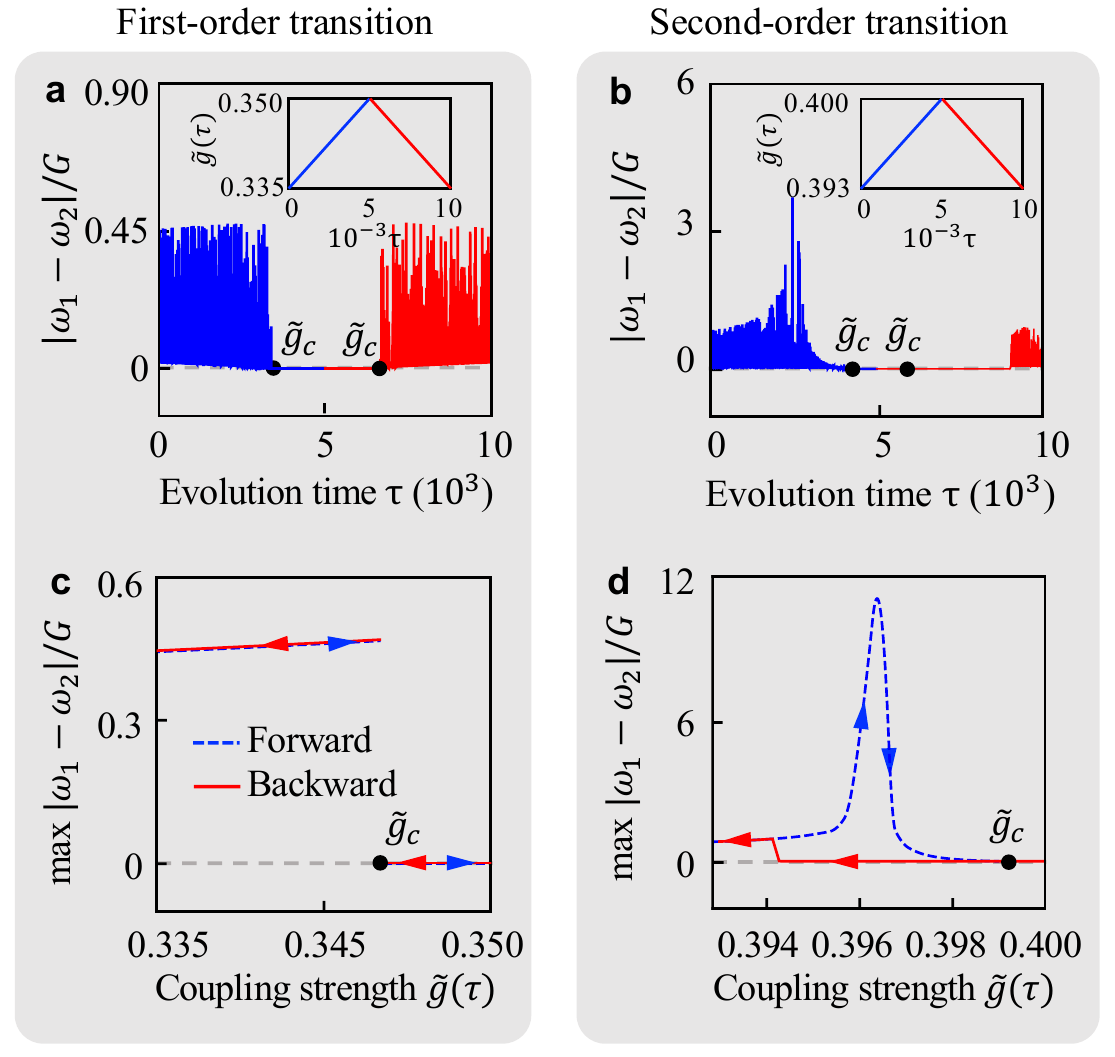}
\caption{\textbf{Hysteresis behavior in frequency difference.}
\textbf{a, b,} Frequency differences $|\omega_1-\omega_2|$ versus the evolution time $\tau$ in the first- and second-order transition regimes. Inset: the real-time evolution of the coupling strength $\tilde g(\tau)$. 
\textbf{c, d,} Maxima of the frequency differences, $\text{max} |\omega_1-\omega_2|$, versus $\tilde g(\tau)$. The plotted parameters are given respectively by the Kerr factor $\tilde \delta=0.1$, and the detuning $\tilde \Delta=0.2$ for (a, c) and $\tilde \Delta=0.3$ for (b, d).  
}
\label{hysteresis}
\end{figure}
\vspace{6pt}

\noindent \textbf{Hysteresis behavior.}
The two types of synchronization transitions present distinct hysteresis behaviors near the critical coupling strength $\tilde g_c$.
In Fig.~\ref{hysteresis}, the frequency differences $|\omega_1-\omega_2|$ and their maxima, $\text{max}|\omega_1-\omega_2|$, are plotted, when the real-time coupling strength $\tilde g(\tau)$ slowly increases (forward) and then decreases (backward).
In the first-order transition regime, whatever the direction $\tilde g(\tau)$ moves, the synchronization emerges or disappears at the same $\tilde g_\text{c}$ as derived in the static analysis [see Fig.~\ref{hysteresis}\textbf{a}].
The maximal frequency differences in the forward and backward evolutions are identical in Fig.~\ref{hysteresis}\textbf{c}, coinciding with Fig.~\ref{surface_g}\textbf{a}.
In the second-order transition regime, although the synchronization also emerges at $\tilde g_\text{c}$ in the forward trip, it does not disappear at the same critical point in the backward trip [see Fig.~\ref{hysteresis}\textbf{b}].
Actually, the synchronization survives far below $\tilde g_\text{c}$, even into the statically inaccessible region in Fig.~\ref{surface_g}\textbf{c}.
Further calculation of the maximal frequency differences reveals a hysteresis loop in this case [see Fig.~\ref{hysteresis}\textbf{d}].
The forward half of the loop remains the same as the curve in Fig.~\ref{surface_g}\textbf{b}, while the backward half is beneath it.
The critical coupling strength under static model does not apply under a dynamical model with the second-order transition.
 As explained in the Supplementary Material \cite{supp}, the $\tilde g(\tau)$ passes $\tilde g_\text{c}$ with the emergent of new non-zero eigenvalues proportional to $-\dot{\tilde g}\mathbf{J}^{-1}$. This mechanism is thus attributed to the singularity of $\mathbf{J}^{-1}$ and the altering direction of real time $\tilde g(\tau)$.
The hysteresis property breaks the minimal coupling required for the synchronization, and the consequent temporal nonreciprocity enables the reading out of coupling history as was done in the ferromagnetic materials \cite{buschow2003physics}. 
\vspace{6pt}

\noindent\textbf{Discussion and Conclusion}

\noindent In summary, we have presented the mode synchronization of two self-sustained optical microresonators which are largely detuned and linearly coupled together. The synchronization is accompanied with a process of spontaneous symmetry breaking, taking the form of the first- and second-order transitions. 
First, when the synchronization takes place, the high transient frequencies of both two modes collapse, offering a possible solution of the frequency mismatch problem in integrating optical microresonators. The phase noise of the coupled system is dramatically reduced, revealing the spontaneous symmetry preservation and paving the way for error-tolerant device fabrication \cite{bagheri2013photonic}. Second, the topological character transitions cast lights on many-body physics. The experimental realization can be approached by coupling two toroid cavities etched with the same mask. Raman gain is applied separately to each cavity at tunable pump frequencies. With additional thermal control of refractive index, the perfect phase matching and adjustable mode frequency difference are achievable. Noting that the state space evolution corresponds to non-trivial degeneration of a ring into a point in our model. During the synchronization of three resonators, however, the transition also includes non-trivial degeneration of the torus into a ring. The multiple-torus topological structure in massively coupled resonators shall offer new insights in many-body physics \cite{armstrong2013basic}.  
Finally, in the second-order transition regime, an unconventional hysteresis behavior has been predicted, breaking the limitation on the critical coupling strength. The coupling history of the resonators can be logged over a short period autonomously, which is desirable in all-optical memory designs \cite{chen2011all,rios2015integrated}.
These results thus show great potential in studying all-optical memory, coupled cavity quantum electrodynamics and many-body optical physics.

\subsection*{Disclosures}
The authors declare that they have no competing financial interests.

\acknowledgments 
We thank Linran Fan, Qi-Tao Cao and Mian Zhang for fruitful discussions. This work was supported by the National Key R\&D Program of China (Grant No. 2016YFA0301302), NSFC (Grant Nos.~11825402, 61435001, 11654003, and 11674200), and High-Performance Computing Platform of Peking University. 



\begin{thebibliography}{10}

\bibitem{van1928heart}
B.~Van Der~Pol and J.~Van Der~Mark, ``Lxxii. the heartbeat considered as a
  relaxation oscillation, and an electrical model of the heart,'' {\em The
  London, Edinburgh, and Dublin Philosophical Magazine and Journal of Science}
  {\bf 6}(38), 763--775  (1928).

\bibitem{fitzhugh1961neuronimpulses}
R.~FitzHugh, ``Impulses and physiological states in theoretical models of nerve
  membrane,'' {\em Biophys. J.} {\bf 1}(6), 445--466  (1961).

\bibitem{buck1968fireflymechanism}
J.~Buck and E.~Buck, ``Mechanism of rhythmic synchronous flashing of fireflies:
  Fireflies of southeast asia may use anticipatory time-measuring in
  synchronizing their flashing,'' {\em Science} {\bf 159}(3821), 1319--1327
  (1968).

\bibitem{huygens1897oeuvres}
C.~Huygens, {\em Oeuvres compl{\`e}tes}, vol.~7, M. Nijhoff  (1897).

\bibitem{oliveira2015huygens}
H.~M. Oliveira and L.~V. Melo, ``Huygens synchronization of two clocks,'' {\em
  Sci. Rep.} {\bf 5}, 11548  (2015).

\bibitem{kurths2001synchronization}
J.~Kurths, A.~Pikovsky, and M.~Rosenblum, {\em Synchronization: a universal
  concept in nonlinear sciences}, Cambridge University Press New York  (2001).

\bibitem{bregni2002clock}
S.~Bregni, {\em Synchronization of digital telecommunications networks},
  vol.~27, Wiley New York  (2002).

\bibitem{bagheri2011dynamic}
M.~Bagheri, M.~Poot, M.~Li, {\em et~al.}, ``Dynamic manipulation of
  nanomechanical resonators in the high-amplitude regime and non-volatile
  mechanical memory operation,'' {\em Nat. Nanotechnol.} {\bf 6}(11), 726
  (2011).

\bibitem{mahboob2008bit}
I.~Mahboob and H.~Yamaguchi, ``Bit storage and bit flip operations in an
  electromechanical oscillator,'' {\em Nat. Nanotechnol.} {\bf 3}(5), 275
  (2008).

\bibitem{hoppensteadt2001synchronization_neurocom}
F.~C. Hoppensteadt and E.~M. Izhikevich, ``Synchronization of mems resonators
  and mechanical neurocomputing,'' {\em IEEE Trans. Circuits Syst. I, Fundam.
  Theory Appl.} {\bf 48}(2), 133--138  (2001).

\bibitem{nijmeijer2003synchronization}
H.~Nijmeijer and A.~Rodriguez-Angeles, {\em Synchronization of mechanical
  systems}, vol.~46, World Scientific  (2003).

\bibitem{shim2007synchronized}
S.-B. Shim, M.~Imboden, and P.~Mohanty, ``Synchronized oscillation in coupled
  nanomechanical oscillators,'' {\em Science} {\bf 316}(5821), 95--99  (2007).

\bibitem{Heinrich2011collective}
G.~Heinrich, M.~Ludwig, J.~Qian, {\em et~al.}, ``Collective dynamics in
  optomechanical arrays,'' {\em Phys. Rev. Lett.} {\bf 107}, 043603  (2011).

\bibitem{holmes2012sync}
C.~A. Holmes, C.~P. Meaney, and G.~J. Milburn, ``Synchronization of many
  nanomechanical resonators coupled via a common cavity field,'' {\em Phys.
  Rev. E} {\bf 85}, 066203  (2012).

\bibitem{zhang2012synchronization}
M.~Zhang, G.~S. Wiederhecker, S.~Manipatruni, {\em et~al.}, ``Synchronization
  of micromechanical oscillators using light,'' {\em Phys. Rev. Lett.} {\bf
  109}(23), 233906  (2012).

\bibitem{zhang2015synchronization}
M.~Zhang, S.~Shah, J.~Cardenas, {\em et~al.}, ``Synchronization and phase noise
  reduction in micromechanical oscillator arrays coupled through light,'' {\em
  Phys. Rev. Lett.} {\bf 115}(16), 163902  (2015).

\bibitem{peano2015topological}
V.~Peano, C.~Brendel, M.~Schmidt, {\em et~al.}, ``Topological phases of sound
  and light,'' {\em Phys. Rev. X} {\bf 5}(3), 031011  (2015).

\bibitem{2015master}
S.~Y. Shah, M.~Zhang, R.~Rand, {\em et~al.}, ``Master-slave locking of
  optomechanical oscillators over a long distance,'' {\em Phys. Rev. Lett.}
  {\bf 114}, 113602  (2015).

\bibitem{bagheri2013photonic}
M.~Bagheri, M.~Poot, L.~Fan, {\em et~al.}, ``Photonic cavity synchronization of
  nanomechanical oscillators,'' {\em Phys. Rev. Lett.} {\bf 111}(21), 213902
  (2013).

\bibitem{li2016oe}
T.~Li, T.-Y. Bao, Y.-L. Zhang, {\em et~al.}, ``Long-distance synchronization of
  unidirectionally cascaded optomechanical systems,'' {\em Opt. Express} {\bf
  24}(11), 12336--12348  (2016).

\bibitem{2017locking}
E.~Gil-Santos, M.~Labousse, C.~Baker, {\em et~al.}, ``Light-mediated cascaded
  locking of multiple nano-optomechanical oscillators,'' {\em Phys. Rev. Lett.}
  {\bf 118}, 063605  (2017).

\bibitem{cross2004synchronization}
M.~Cross, A.~Zumdieck, R.~Lifshitz, {\em et~al.}, ``Synchronization by
  nonlinear frequency pulling,'' {\em Phys. Rev. Lett.} {\bf 93}(22), 224101
  (2004).

\bibitem{Agrawal2013locked}
D.~K. Agrawal, J.~Woodhouse, and A.~A. Seshia, ``Observation of locked phase
  dynamics and enhanced frequency stability in synchronized micromechanical
  oscillators,'' {\em Phys. Rev. Lett.} {\bf 111}, 084101  (2013).

\bibitem{walter2014quantum}
S.~Walter, A.~Nunnenkamp, and C.~Bruder, ``Quantum synchronization of a driven
  self-sustained oscillator,'' {\em Phys. Rev. Lett.} {\bf 112}(9), 094102
  (2014).

\bibitem{pecora2014cluster}
L.~M. Pecora, F.~Sorrentino, A.~M. Hagerstrom, {\em et~al.}, ``Cluster
  synchronization and isolated desynchronization in complex networks with
  symmetries,'' {\em Nat. Commun.} {\bf 5}, 4079  (2014).

\bibitem{matheny2014phase}
M.~H. Matheny, M.~Grau, L.~G. Villanueva, {\em et~al.}, ``Phase synchronization
  of two anharmonic nanomechanical oscillators,'' {\em Phy. Rev. Lett.} {\bf
  112}(1), 014101  (2014).

\bibitem{lorch2016genuine}
N.~L{\"o}rch, E.~Amitai, A.~Nunnenkamp, {\em et~al.}, ``Genuine quantum
  signatures in synchronization of anharmonic self-oscillators,'' {\em Phys.
  Rev. Lett.} {\bf 117}(7), 073601  (2016).

\bibitem{kuramochi2014large}
E.~Kuramochi, K.~Nozaki, A.~Shinya, {\em et~al.}, ``Large-scale integration of
  wavelength-addressable all-optical memories on a photonic crystal chip,''
  {\em Nat. Photon.} {\bf 8}(6), 474  (2014).

\bibitem{zhang2019electronically}
M.~Zhang, C.~Wang, Y.~Hu, {\em et~al.}, ``Electronically programmable photonic
  molecule,'' {\em Nature Photonics} {\bf 13}(1), 36  (2019).

\bibitem{liu2014coherentQED}
Y.-C. Liu, X.~Luan, H.-K. Li, {\em et~al.}, ``Coherent polariton dynamics in
  coupled highly dissipative cavities,'' {\em Phys. Rev. Lett.} {\bf 112}(21),
  213602  (2014).

\bibitem{hwang2016quantum}
M.-J. Hwang and M.~B. Plenio, ``Quantum phase transition in the finite
  jaynes-cummings lattice systems,'' {\em Phys. Rev. Lett.} {\bf 117}(12),
  123602  (2016).

\bibitem{tangpanitanon2016topological}
J.~Tangpanitanon, V.~M. Bastidas, S.~Al-Assam, {\em et~al.}, ``Topological
  pumping of photons in nonlinear resonator arrays,'' {\em Phys. Rev. Lett.}
  {\bf 117}(21), 213603  (2016).

\bibitem{ludwig2013quantum}
M.~Ludwig and F.~Marquardt, ``Quantum many-body dynamics in optomechanical
  arrays,'' {\em Phys. Rev. Lett.} {\bf 111}(7), 073603  (2013).

\bibitem{jang2018synchronization}
J.~K. Jang, A.~Klenner, X.~Ji, {\em et~al.}, ``Synchronization of coupled
  optical microresonators,'' {\em Nat. Photon.} {\bf 12}, 688--693  (2018).

\bibitem{yang2017counter}
Q.-F. Yang, X.~Yi, K.~Y. Yang, {\em et~al.}, ``Counter-propagating solitons in
  microresonators,'' {\em Nat. Photon.} {\bf 11}(9), 560--564  (2017).

\bibitem{jirauschek2006gaussian}
C.~Jirauschek and F.~X. K{\"a}rtner, ``Gaussian pulse dynamics in gain media
  with kerr nonlinearity,'' {\em J. Opt. Soc. Am. B} {\bf 23}(9), 1776--1784
  (2006).

\bibitem{liu2017symmetry}
D.~Liu, B.~Zhen, L.~Ge, {\em et~al.}, ``Symmetry, stability, and computation of
  degenerate lasing modes,'' {\em Phys. Rev. A} {\bf 95}(2), 023835  (2017).

\bibitem{lee2013quantum}
T.~E. Lee and H.~Sadeghpour, ``Quantum synchronization of quantum van der pol
  oscillators with trapped ions,'' {\em Phys. Rev. Lett.} {\bf 111}(23), 234101
   (2013).

\bibitem{walls2007quantum}
D.~F. Walls and G.~J. Milburn, {\em Quantum optics}, Springer Science \&
  Business Media  (2007).

\bibitem{gilles1993two}
L.~Gilles and P.~Knight, ``Two-photon absorption and nonclassical states of
  light,'' {\em Phys. Rev. A} {\bf 48}(2), 1582  (1993).

\bibitem{sebastian2004bifur}
S.~Wieczorek and W.~W. Chow, ``Bifurcations and interacting modes in coupled
  lasers: A strong-coupling theory,'' {\em Phys. Rev. A} {\bf 69}, 033811
  (2004).

\bibitem{wieczorek2004chaos}
S.~Wieczorek and W.~W. Chow, ``Chaos in practically isolated microcavity
  lasers,'' {\em Phys. Rev. Lett.} {\bf 92}(21), 213901  (2004).

\bibitem{carmichael2000statisticaloptics}
H.~Carmichael, {\em Statistical Methods in Quantum Optics 1: Master Equations
  and Fokker-Planck Equations}, Springer  (1999).

\bibitem{supp}
 See Supplementary Material for details.

\bibitem{mari2013measures}
A.~Mari, A.~Farace, N.~Didier, {\em et~al.}, ``Measures of quantum
  synchronization in continuous variable systems,'' {\em Phys. Rev. Lett.} {\bf
  111}(10), 103605  (2013).

\bibitem{buschow2003physics}
K.~H.~J. Buschow and F.~R. Boer, {\em Physics of magnetism and magnetic
  materials}, vol.~92, Springer  (2003).

\bibitem{armstrong2013basic}
M.~A. Armstrong, {\em Basic topology}, Springer Science \& Business Media
  (2013).

\bibitem{chen2011all}
C.-H. Chen, S.~Matsuo, K.~Nozaki, {\em et~al.}, ``All-optical memory based on
  injection-locking bistability in photonic crystal lasers,'' {\em Opt.
  Express} {\bf 19}(4), 3387--3395  (2011).

\bibitem{rios2015integrated}
C.~R{\'\i}os, M.~Stegmaier, P.~Hosseini, {\em et~al.}, ``Integrated
  all-photonic non-volatile multi-level memory,'' {\em Nat. Photon.} {\bf
  9}(11), 725  (2015).

\end{thebibliography}

\bibliographystyle{spiejour}   %

\end{spacing}
\end{document}